\documentclass[prb,aps,preprint,superscriptaddress,floatfix]{revtex4}
\usepackage[latin1]{inputenc}
\usepackage{graphicx}
\usepackage{float}
\usepackage{amsmath}
\usepackage{natbib}
\usepackage{esint}
\usepackage{eucal}
\usepackage{color}
\usepackage{caption}
\usepackage{subfig}

\begin{document}

\title{Ultrathin acoustic parity-time symmetric metasurface cloak}

\author{Hao-xiang Li}
\affiliation{Key Laboratory of Modern Acoustics, MOE, Institute of Acoustics, Department of Physics, Collaborative Innovation Center of Advanced Microstructures, Nanjing University, Nanjing, 210093, China}
\author{Maria Rosendo-L{\'o}pez}
\affiliation{Department of Physics, Universidad Carlos III de Madrid, ES-28916 Legan\`es, Madrid, Spain}
\author{Yi-fan Zhu}
\affiliation{Key Laboratory of Modern Acoustics, MOE, Institute of Acoustics, Department of Physics, Collaborative Innovation Center of Advanced Microstructures, Nanjing University, Nanjing, 210093, China}
\author{Xu-dong Fan}
\affiliation{Key Laboratory of Modern Acoustics, MOE, Institute of Acoustics, Department of Physics, Collaborative Innovation Center of Advanced Microstructures, Nanjing University, Nanjing, 210093, China}
\author{Daniel Torrent}
\affiliation{GROC, UJI, Institut de Noves Tecnologies de la Imatge (INIT), Universitat Jaume I, 12080 Castell\`e, Spain}
\author{Bin Liang}
\affiliation{Key Laboratory of Modern Acoustics, MOE, Institute of Acoustics, Department of Physics, Collaborative Innovation Center of Advanced Microstructures, Nanjing University, Nanjing, 210093, China}
\author{Jian-chun Cheng}
\affiliation{Key Laboratory of Modern Acoustics, MOE, Institute of Acoustics, Department of Physics, Collaborative Innovation Center of Advanced Microstructures, Nanjing University, Nanjing, 210093, China}
\author{Johan Christensen}
\affiliation{Department of Physics, Universidad Carlos III de Madrid, ES-28916 Legan\`es, Madrid, Spain}

\maketitle

\textbf{Invisibility or unhearability cloaks have been made possible by using metamaterials making light or sound flow around obstacle without the trace of reflections or shadows. Metamaterials are known for being flexible building units that can mimic a host of unusual and extreme material responses, which are essential when engineering artificial material properties to realize a coordinate transforming cloak. Bending and stretching the coordinate grid in space requires stringent material parameters, therefore, small inaccuracies and inevitable material losses become sources for unwanted scattering that are decremental to the desired effect. These obstacles further limit the possibility to achieve a robust concealment of sizeable objects from either radar or sonar detection. By using a elaborate arrangement of gain and lossy acoustic media respecting parity-time symmetry, we built an one-way unhearability cloak capable to hide objects seven times larger than acoustic wavelength. Generally speaking, our approach has no limits in terms of working frequency, shape, or size, specifically though, we demonstrate how, in principle, an object of the size of a human can be hidden from audible sound.}

\section{Introduction}

For centuries people have dreamt of an invisibility cloak that can make someone indiscernible for the nacked eye when hidden underneath it. An ideal cloak would involve the suppression of back reflected light to render an object camouflaged, but the shadow behind it must also diminish for truly being able to make an object disappear. More than a decade ago an approach was brought forward based on transformation optics permitting the path of light to be bent around objects to be hidden \cite{Pendry1780,Schurig977,PhysRevLett.101.203901,ValentineCarpet,Ergin337,Ni1310,PhysRevB.80.245115}. This transformational approach to engineer space has ever since been extended to other areas of wave physics for cloaks of unhearability comprising sound and mechanical vibrations \cite{CummerSchurig,TorrentNJP,DehesaScattC,CummerNatMater,StengerPRL,AcReview}. Despite those advances, it remains a fundamental challenge to create an unhearability cloak of ultrathin layer width involving minuscule loss-free materials. Here we propose theoretically and demonstrate experimentally that a parity-time symmetric metasurface incorporating acoustic gain and loss can act as such cloak when insonified from one direction. In contrast to transformation acoustics that implies a coordinate-transformation-based deformation of sound through the accurate distribution of the material properties, our approach to perfectly absorb incoming sound and to re-emit it behind the hidden object solely implies the engineering of a complex acoustic metasurface-impedance. The use of parity-time symmetry enables unique cloaking properties useful in the audible range but also applicable to hide submarines from sonar detection.\\

Non-Hermitian systems that respect parity-time (PT)-symmetry have recently become an active frontier in wave physics due to unprecedented possibilities in guiding both sound and light \cite{Bender1,Bender2}. Most notably, designing complex eigenstates through appropriate balancing of gain and loss provides an unexpected paradigm for exploring non-Hermitian wave control in flourishing areas such as, waveguiding, sensing, communication, and topological insulators \cite{PhysRevLett.100.103904,PhysRevLett.103.093902,NatPhysRuter,NatPhysPeng,MePRL,ZhangNatComm,ABN,PhysRevX.8.031079}. In this work we demonstrate that a sizeable acoustically rigid obstacle (7 times larger than the wavelength) appears hidden to sound waves when coated by a PT symmetric metasurface due to the cancellation of reflections and re-radiation of the impinging field to the far-side. When adjusting the gain-loss contrast to a point where this scenario is reached, unidirectional invisibility (unhearability) is obtained, an effect commonly know as the anisotropic transmission resonance (ATR) \cite{,PRL_invis,ATR_PRA}. \\

Designing PT symmetry is based on manipulating absorption using judicious structures with gain regions and vice versa. The Hamiltonian commutes with the combined PT operator when loss and gain are equally balanced giving rise to entirely real eigenmode frequencies representing the unbroken or exact phase. When the loss and gain contrast exceeds a certain threshold to reach the broken phase, one of the complex eigenmodes exhibits exponential growth while the other one decays exponentially. The transition between these two phases is the non-Hermitian singularity, also known as an exceptional point (EP) where the modes coalesce. The ATR is associated to the flux-conservation process leading to full transparency, i.e., unity transmittance $T=1$, but one-sided reflectionless wave propagation. A special case of the ATR is the unidirectional invisibility phenomenon that not only fulfills the condition of full transmission and vanishing of the reflection from either left $R_L$ or right $R_R$ incidence, but also implies a zero transmission phase signifying the apparent absence of an obstacle to be heard or seen \cite{ATR_PRA}. Several studies, both in the fields of optics and acoustics, have already investigated unidirectional invisibility in one dimensional PT symmetric structures enabling shadow-free acoustic sensors and Bragg-scattering suppression in photonic lattices \cite{RegengsburgNature,AluNatComm}. Here we demonstrate the ability to acoustically cloak a rigid obstacle by covering it by an ultrathin PT metasurface as has been previously simulated for microwave radiation \cite{PhysRevAppliedAlu}. The approach consists in camouflaging the portion of the insonified metasurface through absorption and providing the time-reversed image, i.e., acoustic gain, to the shadow region behind the rigid obstacle (Fig. 1). \\  

\section{Results}
\textbf{Complete absorption of sound}\\
We begin the study by designing the insonified portion of the cloak, whereas the the gain portion will be treated afterwards. Hence, in order to engineer complete scattering cancellation to an incoming plane wave we cover the rigid cylinder of radius $a$ by a lossy metasurface as shown in Fig. 1. The aim is to impedance match the metasurface to the surrounding air to totally absorb the incident wave without reflection. Under the assumption that the air gap separating the ultrathin cloak from the rigid cylinder is substantially smaller than the wavelength of the sound wave, i.e., $k(b-a)\ll 1$ we can write down the necessary complex surface impedance to fulfill camouflaging of the obstacle (see supplementary information for derivations):
\begin{equation}
\begin{aligned}
\text{Re}(Z_s)&\approx Z_0\frac{b}{a}\;\frac{1}{\cos\theta}\\
\text{Im}(Z_s)&\approx Z_0\frac{b}{a}\;k(b-a),
\end{aligned}
\end{equation}
where $\theta$ is the angular position in polar coordinates and $k=2\pi/\lambda$ where $\lambda$ is the wavelength. Eq. (1) states that an impedance match of the metasurface to its surrounding with respect the geometrical parameters $a$ and $b$ is essential in order to achieve complete acoustic energy absorption. In Eq. (1), the free space impedance $Z_0=\rho_0c_0$ where $\rho_0$ and $c_0$ are the mass density and speed of sound in air, respectively. There is a plethora of passive and active metamaterials solutions available capable of complete sound absorption \cite{AcReview}. For the realization of a one-sided compact and lightweight unhearability cloak we use Helmholtz resonators that can be fabricated to absorb sound at broad spectral windows. In the present case, we focus on the audible range and therefore begin the design by engineering sound absorption at a frequency of f = 3 KHz although the approach could be readily realized at other desired frequencies. Strong air oscillations in the neck of these resonators in the presence of viscous losses are responsible for efficient energy dissipation. Conclusively, most absorbed acoustic energy is localized at the neck region, therefore, in order to fully camouflage an object, we pattern the rigid obstacle of radius $a =$ 40 cm by Helmholtz resonators (Fig. 2(a)) and adjust the individual neck parameters $w$ and $t$ accordingly (See supplementary information for their values) to account for the angular variation. By computing the averaged acoustic pressure $\langle p\rangle$ and velocity normal to the resonator surface $\langle v_{\bot}\rangle$ we are able to determine the impedance of the metasurface $Z_s=\langle p\rangle/\langle v_{\bot}\rangle$. In the absence of acoustic backscattering, we predict total absorption of a plane wave at each individual Helmholtz resonator as displayed In Fig. 2(b) via their specific angular position. Correspondingly, we are able to explain full acoustic absorption via surface impedance matching as predicted in Eq. (1) where the real part of the relative metasurface impedance scales according to $1/\cos(\theta)$ and its imaginary counterpart approaches zero for a vanishing gap separation $a \approx b$.\\

\textbf{Acoustic gain adjustment}\\
Perfect absorption removes acoustic backscattering and is the first ingredient of a PT symmetric system. The time-reversed image of this response constitutes acoustic amplification that we implement with an active electric circuit to control an semicircular array of loudspeakers (Fig. 3(a)) \cite{ZhangNatComm,AluNatComm}. In order to implement sound amplification of equal but opposite strength to the absorbing counterpart we must ensure to meet the condition of the ATR that dictates unidirectional-zero reflection at full transmission. Hence, beyond the need of balancing out the acoustic attenuation at the loss semi-shell with the gain counterpart that is expressed through the PT symmetry of the entire metasurface cloak: $Z_s(\theta)=-Z^{*}_s(\pi-\theta)$, we must ensure that the acoustic intensity profile is spatially symmetric. Thus,at the ATR it can be shown that
\begin{equation}
\begin{aligned}
I(\theta)=I(\pi-\theta),
\end{aligned}
\end{equation}
which signifies that when the PT symmetric metasurface cloak is irradiated at the loss portion, the acoustic intensity in the nearest vicinity of an individual Helmholtz resonator (located at $(\pi-\theta)$) equals the intensity at the exact opposite active loudspeaker (located at $\theta$ with respect to Fig. 3(a)) \cite{ATR_PRA}. This property accompanying the ATR condition is extremely useful when adjusting the individual loudspeakers to realize a one-way unhearability cloak. First, as detailed in the method section and the supplementary information, we placed the two jointed semi-shells surrounding the rigid obstacle inside an acoustic waveguide whose rigid walls are covered by absorbing cotton. The acoustic source is formed by an array of loudspeakers that generate plane waves with frequency f = 3 KHz. In order to emit signals from the gain semi-shell perfectly synchronized in phase and amplitude with the impinging signal, a microphone measures the incoming sound field in front of the Helmholtz resonators whose phase and amplitude is processed through a phase shifting and amplifier circuit. The adjustment is performed in relation to the discrete and opposite locations of the Helmholtz resonator and loudspeaker couples whose intensity relation at the ATR is shown in Fig. 3(b). Due to size limitations of the source and the geometrical restrictions of the waveguide, our detection range exhibit an unitary intensity relation upto $\pm 48^{\circ}$ beyond which deviations start to grow. In other words, the subsequent cloaking experiment will be conclusively limited to this range.\\

\textbf{PT symmetry cloak}\\
Fig. 4(a) displays the metasurface cloak in its entirety comprising the jointed non-Hermitian semi-shells surrounding the acrylic obstacle of diameter 80 cm. At an operation frequency of 3 KHz corresponding to an acoustic wavelength of 11 cm, simulations and experimental measurements display how the pressure waves impinging the undecorated acrylic obstacle back-scatter at the side of irradiation but leave and almost soundless shadow at the obstacle's far-side (Fig. 4(b)). Contrary to this, when sound irradiates the non-Hermitian semi-shells that have been tuned to fulfill the aforementioned PT symmetry ATR condition, the acrylic obstacle, whose diameter  is about 7 times larger than the acoustic wavelength, is acoustically camouflaged to match its surrounding via complete absorption, but, more importantly, the acoustic shadow gets eliminated through reconstruction of the impinging wave (Fig. 4(c)). The experimental measurements show that within the test areas, both in front and behind the obstacle, almost perfect plane waves have been sustained rendering the object to be hidden perfectly unhearable and concealed. \\

\textbf{Discussion}\\
Further improvement can be achieved by enlarging the loudspeaker array to launch a near-ideal plane wave. Also, decorating an object of arbitrary shape with gain and loss units in response to a point source or more complicated wave shapes greatly broadens the usage of PT symmetry based acoustic cloaks. We implemented the proof of concept by means of Helmholtz resonators to suppress back reflected sound via resonant absorption. In analogy to the implemented active gain component, active loss control would enrich the possibility to eliminate back-scattering and to provide an acoustic camouflage dynamically at a wider spectral range. Extensions toward an acoustical concealment of three dimensional bodies by ultrathin non-Hermitian shells is another avenue worth pursuing. \\
In conclusion, We have derived a theoretical recipe to realize an acoustic unhearability cloak via PT symmetry. By combining loss and gain structures, we showed that reflected sound is eliminated from an insonified body to be concealed and how it is reconstructed at the rear side of it via the anisotropic transmission resonance. Full wave simulations and measurement data support the theoretical predictions in creating a cloak based on a single but non-Hermitian shell structure.

\section{Acknowledgements}
J. C. acknowledges the support from the European Research Council (ERC) through the Starting Grant No. 714577 PHONOMETA and from the MINECO through a Ram\'on y Cajal grant (Grant No. RYC-2015-17156).


\newpage
\section{Bibliography}
\bibliographystyle{naturemag}

\newpage

\section{Methods}

\textbf{Numerical simulations}. The full-wave simulations presented in the paper are performed using the finite-element solver Comsol Multiphysics. We employed the acoustic module comprising viscothermal losses and modelled the Helmholtz resonators as acoustically rigid materials. The surface impedance is computed via the averaged acoustic pressure and normal velocity $Z_s=\langle p\rangle/\langle v_{\bot}\rangle$ at the resonator surface. Acoustic gain has been imposed at the respective boundary through amplifying outgoing wave.\\

\textbf{Fabrication of the cloak}. The obstacle to be cloaked was realized via a cylindrical acrylic shell with an outer diameter of 80 cm. The Helmholtz resonators responsible for the loss components and the active loudspeakers providing acoustic gain were mounted onto the shell. The individual Helmholtz resonators were 3D printed with thermoplastics whose geometrical parameters are tabulated in the supplementary information.\\

\textbf{Measurements}. The experiment is carried out in a two dimensional waveguide with a uniform height of 6 cm. As can be seen in the illustration below,
\begin{figure}[htbp]
\addtocounter{figure}{-1}
\begin{center}
\includegraphics[width=0.8\columnwidth]{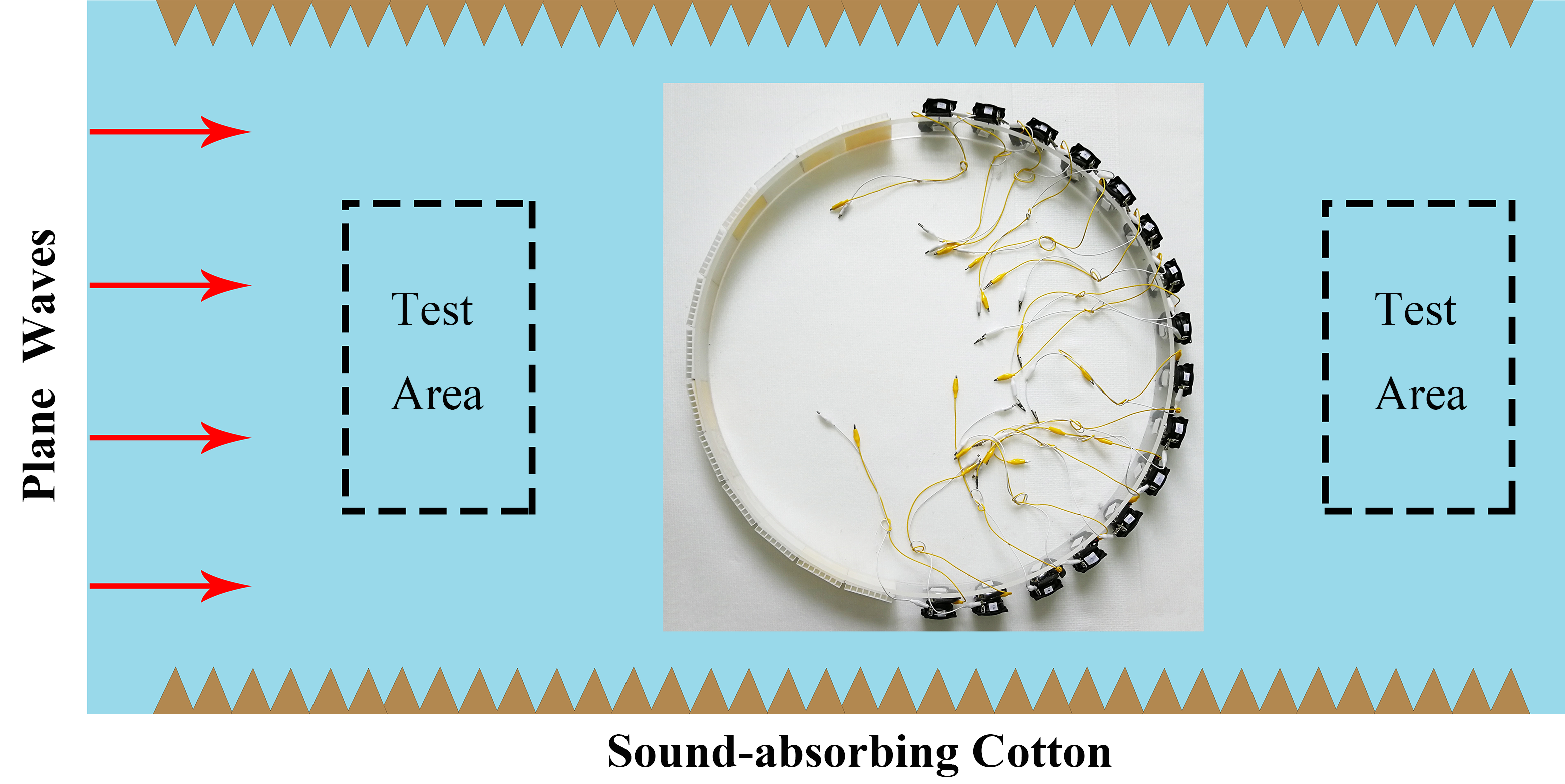}
\caption{Experimental implementation of the parity-time symmetric metasurface cloak.}
\end{center}
\end{figure}
a line speaker array was employed to launch an incoming plane wave. In order to reduce unwanted reflections we covered the inner walls of the waveguide with sound-absorbing cotton. The experimental implementation is detailed in the supplementary information.

\newpage

\section{Figures}

\begin{figure}[htbp]
\begin{center}
\includegraphics[width=0.7\columnwidth]{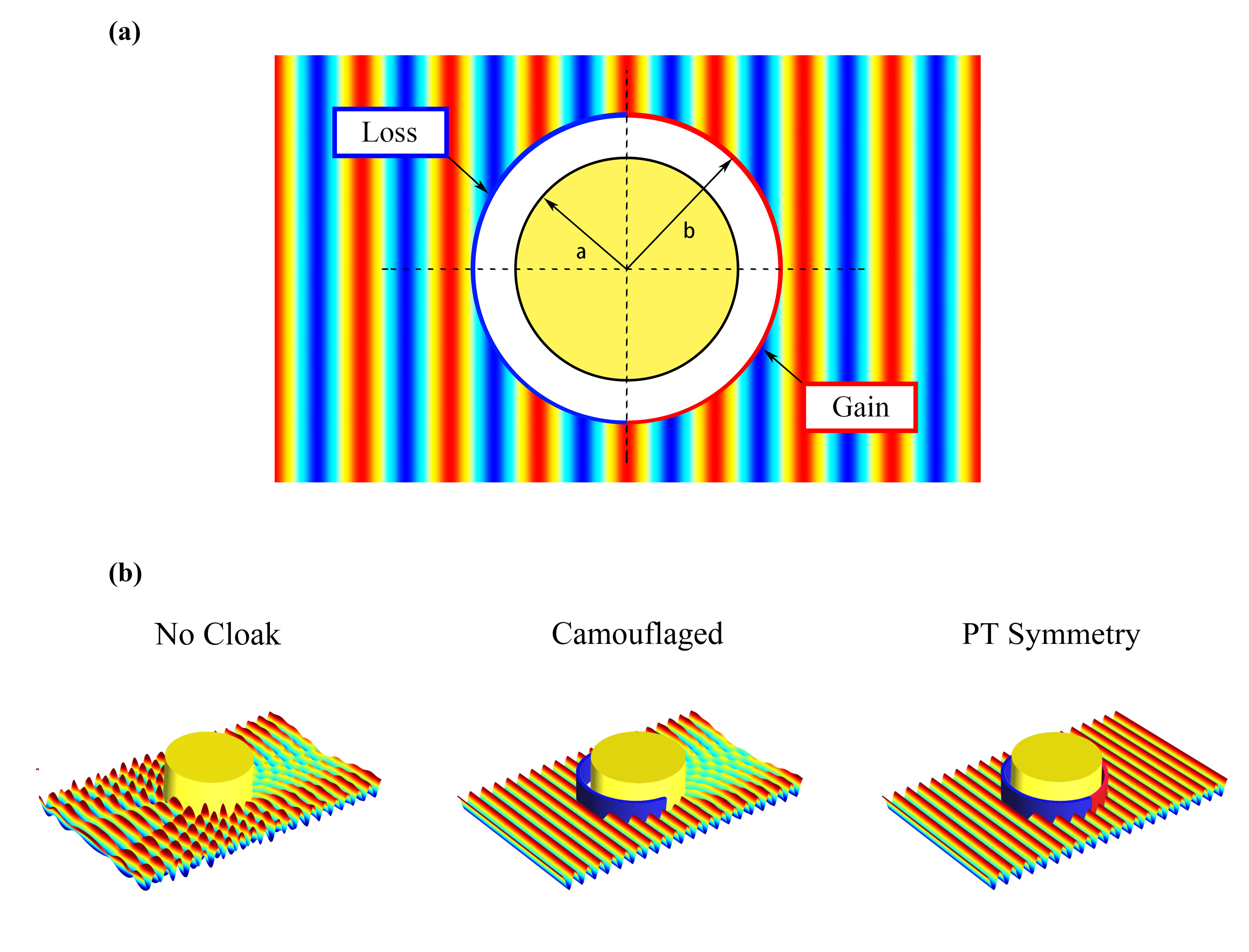}
\caption{(a) A rigid cylinder is covered by an ultrathin PT symmetric metasurface. The left (right) semicircle metasurface contain acoustic loss (gain) to fully absorb incoming (re-emit outgoing) sound waves. (b) Three scenarios are exemplified: No cloak, comprising strong back-scattering and shadow; camouflaging through complete absorption with a lossy semicircle metasurface only; cloaking via PT symmetry.}
\end{center}
\end{figure}

\begin{figure}[htbp]
\begin{center}
\includegraphics[width=0.8\columnwidth]{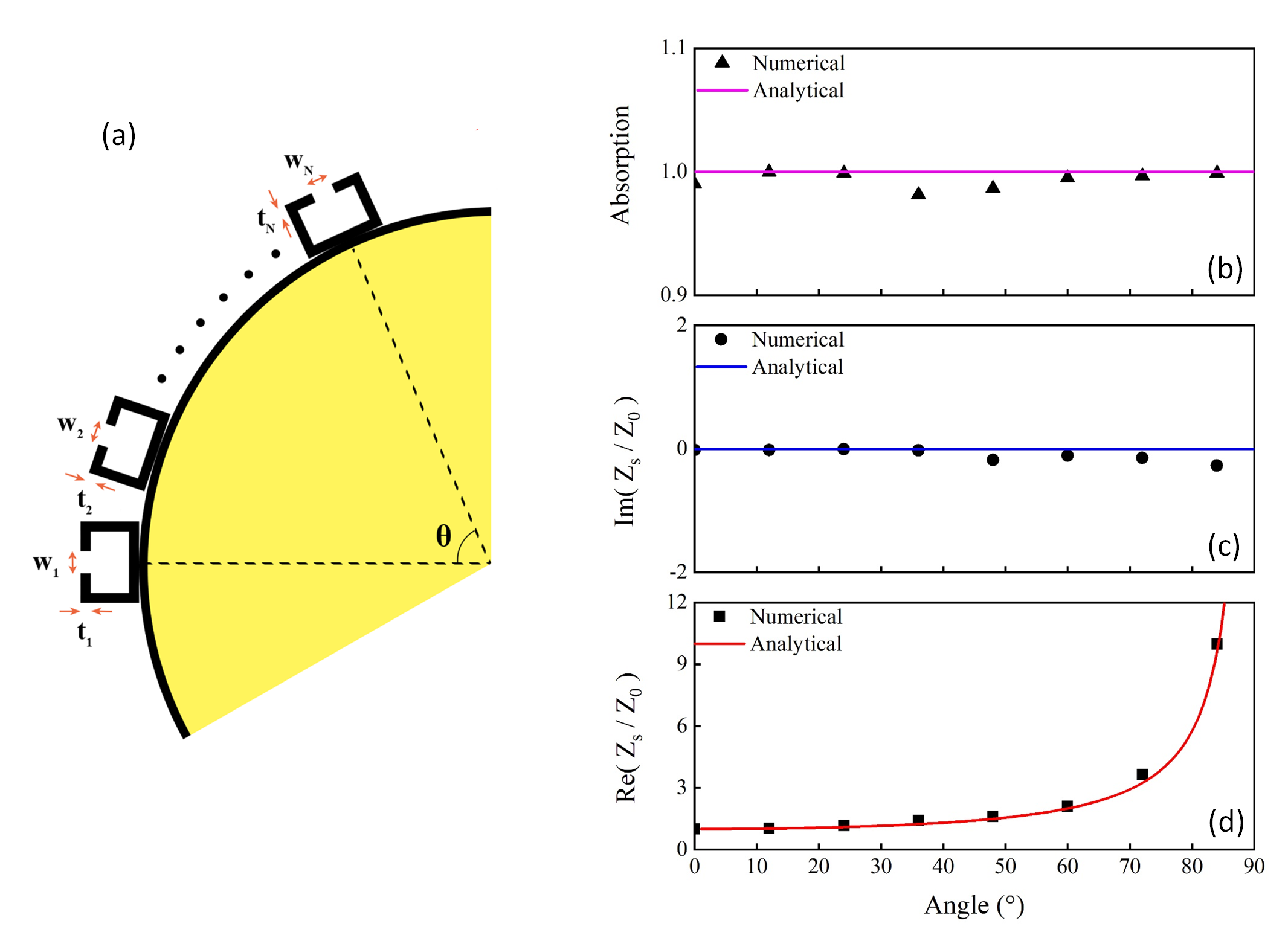}
\caption{(a) The insonified portion of the cloak ($|\theta|<\pi/2$) is patterned by sound absorbing Helmholtz resonators. The angular dependence of their resonances has been tuned via the resonator neck width $w$ and depth $t$. (b-d) The theoretically and numerically computed absorption and impedance match is presented in dependence to the angle $\theta$, i.e, the position of the individual Helmholtz resonators.}
\end{center}
\end{figure}

\begin{figure}[htbp]
\begin{center}
\includegraphics[width=0.8\columnwidth]{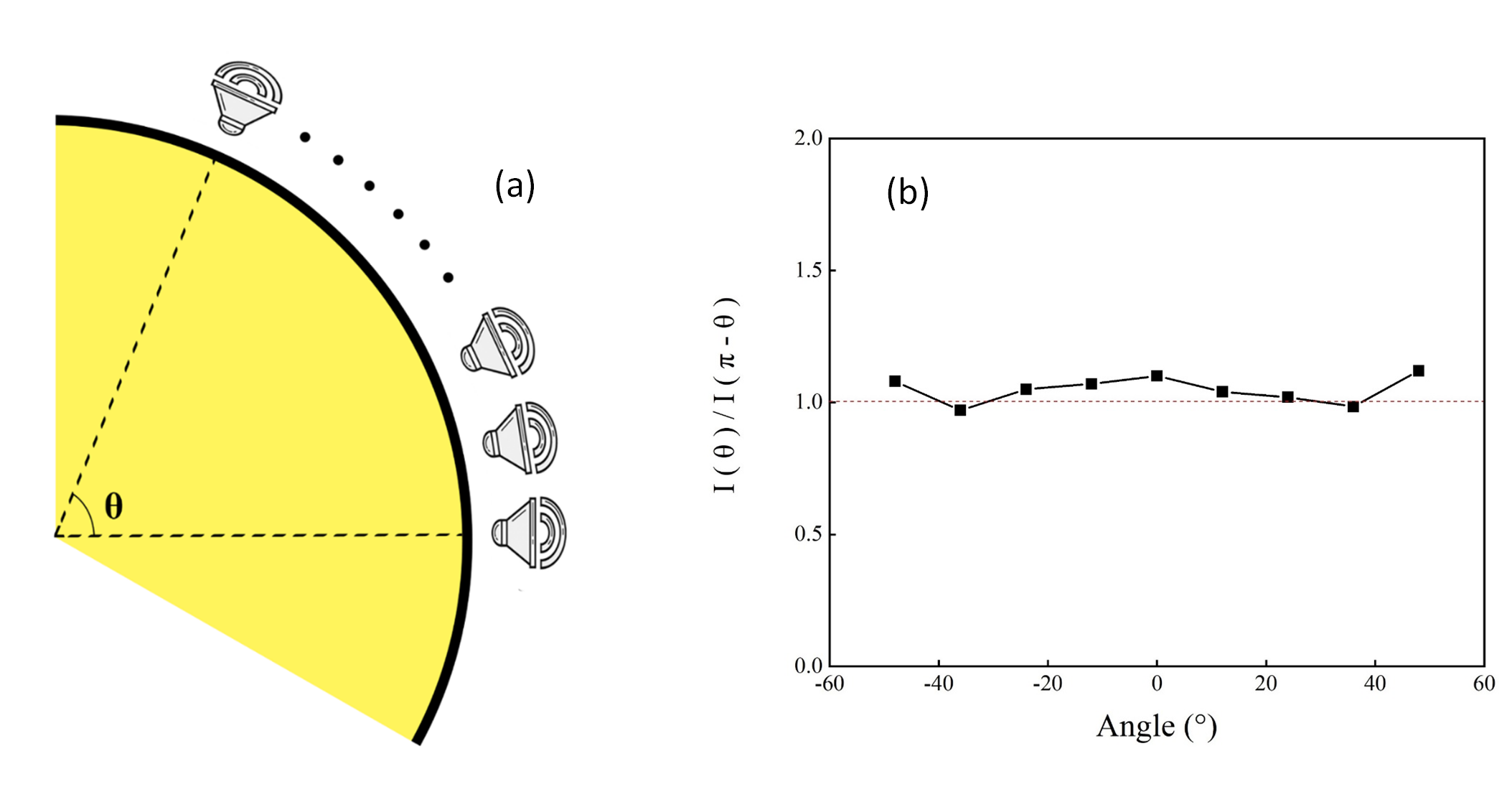}
\caption{(a) The amplifying portion of the metasurface cloak has been decorated by loudspeakers that are controlled by gain circuits. (b) The anisotropic transmission resonance with omnidirectional full transmission and one-sided zero reflection, is accompanied by a spatial symmetry of the measured intensity profile $I(\theta)=I(\pi-\theta)$, which was the experimental parameter for the gain adjustment.}
\end{center}
\end{figure}

\begin{figure}[htbp]
\begin{center}
\includegraphics[width=1\columnwidth]{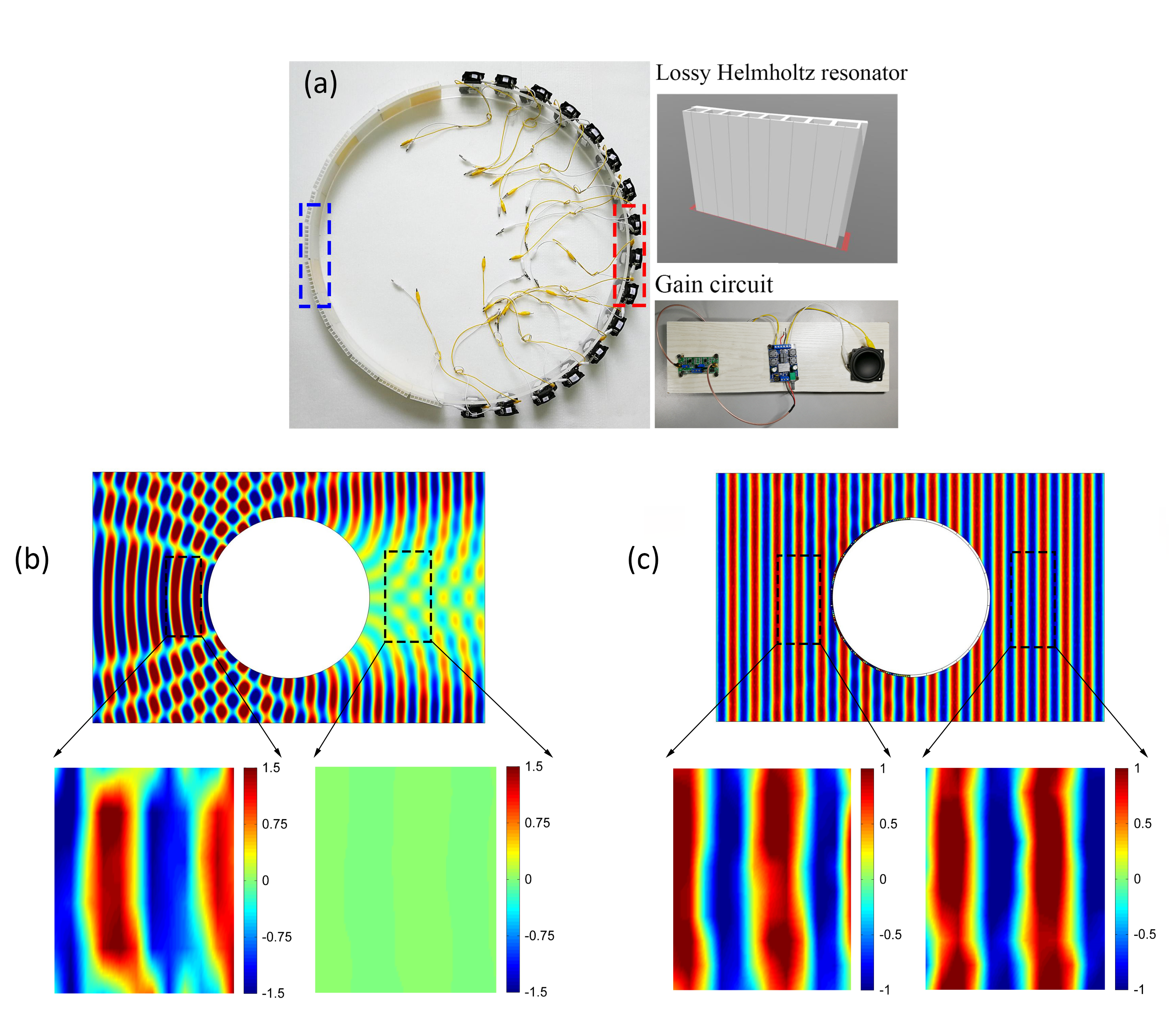}
\caption{(a) Experimental realization of the PT symmetric metasurface cloak made out of two jointed semi-shells (radius $a=$ 40 cm): (a) lossy Helmholtz resonator array and active loudspeakers controlled by gain circuits. (b) Full wave simulations of the pressure field of a bare rigid cylinder when insonified from the left by a plane wave at f = 3 kHz. The dashed test areas have been experimentally measured. (c) Simulations of the metasurface cloak surrounding the rigid obstacle and the corresponding measurements at the front and the backside of the decorated obstacle.}
\end{center}
\end{figure}
\end{document}